# Route toward high-speed nano-magnonics

B. Divinskiy[1*], V. E. Demidov[1], S. O. Demokritov[1,2], A. B. Rinkevich[2], and S. Urazhdin[3]

[1]*Institute for Applied Physics and Center for Nonlinear Science, University of Muenster, Corrensstrasse 2-4, 48149 Muenster, Germany*

[2]*Institute of Metal Physics, Ural Division of RAS, Yekaterinburg 620041, Russia*

[3]*Department of Physics, Emory University, Atlanta, GA 30322, USA*

We study experimentally the possibility to utilize pulses of pure spin current, produced via the nonlocal spin injection mechanism, to generate short packets of spin waves propagating in nanoscale magnetic waveguides. The spatially and time-resolved micro-focus Brillouin light scattering spectroscopy measurements demonstrate that the excitation by spin current results in extremely fast transient response, enabling efficient generation of short spin-wave packets with duration down to a few nanoseconds. The proposed method opens a route for the implementation of high-speed magnonic systems for transmission and processing of information on the nanoscale.

The emerging field of magnonics[1-4] explores the mechanisms enabling transmission and processing of information by spin waves. The rapidly growing interest in this field can be attributed to a number of important advantages of spin waves as a nanoscale signal carrier. The unique controllability of spin wave characteristics by the magnetic field allows one to efficiently tune their phase, wavelength, and the propagation path[5-10], while the possibility to achieve sub-micrometer wavelength at microwave frequencies[11] makes these waves attractive for the implementation of nanoscale devices.

Downscaling of magnonic devices poses a number of new challenges. In particular, the traditional inductive method for spin wave excitation becomes inefficient at nanoscale due to the increasing requirements for the power density, the unavoidable limitations imposed on the wavelength of the excited waves by the geometry of the inductive antennae, and the difficulty of the impedance matching of the latter. Although these shortcomings have been partially overcome in the recent years[11-14], the inductive excitation is still far from becoming viable for real-world nanoscale applications.

An alternative approach to the excitation of spin waves can be based on the spin-transfer torque (STT) mechanism, which provides the ability to directly convert dc electrical currents into microwave-frequency spin waves on nanoscale[3,15-18]. Unfortunately, the limitations of the geometry of traditional STT devices operating with spin-polarized electric currents place significant constraints on their compatibility with magnonic devices. By contrast, a significant geometric flexibility can be achieved in the STT devices operated by pure spin currents generated either by the spin Hall effect or by the nonlocal spin injection (NLSI)[19,20]. Indeed, excitation of propagating spin waves by pure spin currents generated by the NLSI mechanism has been recently demonstrated,[18] providing a simple and flexible route compatible with magnonic devices.

The ability of a spin-wave source to generate short wave packets is particularly important for the practical implementations of high-speed integrated magnonic circuits. The performance of the traditional inductive excitation technique is very limited in this respect, since the externally generated microwave signal has to be pulse-modulated by semiconductor switches, which are generally characterized by a relatively low power efficiency, and on-off times of at least several nanoseconds. The fastest spin-wave excitation rate demonstrated so far was achieved by utilizing ultra-short laser pulses[21-23]. However, this approach requires a high-power femtosecond optical source, and therefore has significant technological limitations.

In this Letter, we present a study of the temporal characteristics of the NLSI-based mechanism for the excitation of propagating spin waves. We show that this mechanism is sufficiently fast to enable generation of short spin-wave packets with the duration down to 2 ns, close to the best results achieved by using optical-pulse excitation[21-23]. Moreover, we find that the intense spin-wave packets generated by the pure spin current experience a nonlinear compression while propagating in the magnonic nano-waveguides, which further reduces their temporal width. A similar mechanism is responsible for the formation of nonlinear spin-wave solitons. It allows one to compensate for the dispersive broadening of spin-wave packets by engineering the nonlinear characteristics of magnonic transmission lines, resulting in improved information flow capacity.

Figure 1(a) shows the layout of the studied device. It consists of a vertical NLSI spin valve[24,25] dynamically coupled to a spin-wave waveguide. The spin valve is formed by a circular 60-nm diameter Au nanocontact underneath an extended CoFe(8 nm)/Cu(20nm)/Py(5 nm) trilayer, while the waveguide is formed by a 20 nm-thick and 500 nm-wide Py strip fabricated on top of the trilayer, and terminated at the distance of 150 nm from the nanocontact. The waveguiding strip is tapered to the width of 300 nm at the edge facing the nanocontact. This geometry compensates the effects of the inhomogeneous dipolar magnetic field at the edge, providing a uniform distribution of the internal static magnetic field throughout the entire waveguide. The driving electrical current is injected into the system through the nanocontact, and drained through the trilayer toward the side electrode, as shown by the arrow in Fig. 1(a). The arrow in the inset in Fig. 1(a) shows the corresponding flow of electrons. The injected electrons become spin-polarized due to the spin-dependent scattering in CoFe and at the Cu/CoFe interface[26], resulting in spin accumulation in Cu above the nanocontact. Spin diffusion away from this region produces a spin current flowing into the Py layer, exerting STT on its magnetization. The magnetizations of both the CoFe and the Py layers are aligned with the saturating static in-plane magnetic field $H_0$ oriented perpendicular to the waveguide axis. For positive driving electric currents, as defined by the arrow in Fig. 1(a), the magnetic moment carried by the spin current is antiparallel to the magnetization of the Py layer, resulting in the STT compensating the dynamic magnetic damping. When damping is completely compensated by the spin current, the magnetization of the Py layer exhibits microwave-frequency auto-oscillations in the area above the nanocontact[25].

We detect the current-induced magnetization dynamics by using the micro-focus Brillouin light scattering (BLS)[8,27] spectroscopy. We focus the probing laser light on the surface of the Py film into a diffraction-limited spot (Fig. 1(a)), and analyze the spectrum of light inelastically scattered from the magnetic oscillations. The BLS signal is proportional to the intensity of magnetization oscillations at a given frequency. The technique provides information about the dynamic magnetization with simultaneous temporal, spectral and spatial resolution.

Figure 1(b) shows the BLS spectra of current-induced auto-oscillations measured by positioning the probing laser spot at the location of the nanocontact. The onset of auto-oscillations is signified by the emergence of an intense spectral peak at currents above $I_C$ = 3.6 mA. The intensity of the auto-oscillation peak gradually increases with increasing current up to 7 mA, and starts to decrease at larger currents. The central frequency monotonically decreases with increasing *I*, due to the nonlinear frequency shift[28,29]. Note that the linewidth of the spectral peaks in Fig. 1(b) is determined by the limited frequency resolution of the BLS technique. High-resolution electronic measurements performed in Ref. 30 showed that the actual room-temperature linewidth of the auto-oscillations in the NLSI devices is a few megahertz, making the oscillators suitable for magnonic applications requiring a high level of spin wave coherence.

The auto-oscillation frequency of the NLSI oscillator is always smaller than the frequency $f_0$ of the uniform ferromagnetic resonance (FMR) in the extended Py film, which is shown in Fig. 1(b) with a dashed line. The value of $f_0$ was determined from independent BLS measurements of thermal magnetization fluctuations at $I = 0$. Since the extended Py film does not support propagating spin-wave states at the frequency of the auto-oscillations, the latter are spatially localized and do not radiate spin waves into the surrounding extended film. The demagnetization effects in the profiled waveguide result in the downshift of the spin wave spectrum relative to the extended Py film[31]. As a consequence, the frequencies of the propagating waveguide modes are well matched with those of the auto-oscillations, enabling efficient spin-wave emission into the waveguide due to the auto-oscillation[18].

To determine the temporal characteristics of the NLSI mechanism, we applied short square pulses of current to the nanocontact, while performing time-resolved BLS measurements synchronized with the pulses. Figure 2 illustrates the results obtained by using 100 ns-long pulses of current with the repetition period of 500 ns, with the BLS spot centered on the nanocontact. The intensity of the dynamic magnetization exhibits a rapid increase within a few nanoseconds after the start of the driving-current pulse at *t*=0, reaches a broad plateau, and finally abruptly drops after the driving current is switched off, as shown in Fig. 2(a) for two different amplitudes of the current pulse. Note that the rate of the intensity growth at the beginning of the pulse increases with increasing $I$, in agreement with the theory of STT-driven dynamics[32]. Indeed, the effects of STT at $I>I_C$ can be described as negative effective damping, whose magnitude determines the rate of the exponential increase of the oscillation amplitude with time. The negative damping rate increases with increasing $I$, resulting in a faster amplitude increase.

To quantitatively characterize the transient dynamics of the NLSI oscillator, we fit the leading edge of the auto-oscillation intensity pulse by the exponential function $A*\exp(t/\tau)$, as shown in Fig. 2(b), and determine the rise time $\tau$. The latter is defined as the time required for the intensity of the oscillations to increase by a factor of e. As shown in Fig. 2(c), $\tau$ rapidly decreases with increasing current, reaching values below 1 ns at $I > 4.2$ mA. At $I > 8$ mA, the measured rise time saturates at about 0.45 ns, which is close to the limit of the temporal resolution of the measurement. These results clearly indicate that the NLSI-based STT mechanism is sufficiently fast for the generation of nanosecond-scale pulses of dynamic magnetization.

Next, we characterize the efficiency of the dynamical excitation by the spin current as a function of the pulse duration. We fix the current amplitude at 7 mA, and vary the temporal width $w_d$ of the driving current pulses. Figure 3(a) shows the dynamic response of the NLSI oscillator for three different driving pulse widths. As seen from these data, the generated pulse of the dynamic magnetization maintains a nearly rectangular shape and a constant peak intensity for $w_d$ down to 10 ns. As $w_d$ is reduced to 3 ns, the dynamic-magnetization pulse becomes almost Gaussian-shaped, and its peak intensity reduces by about 40 %, consistent with the expected effects of the finite response time of the NLSI oscillator. The dependence of the peak oscillation intensity on the width of the driving-current pulse (Fig. 3(b)), determined at a constant amplitude $I$=7 mA, shows that the efficiency of the driving mechanism rapidly diminishes at $w_d$ < 3 ns. In particular, at $w_d$ < 2.5 the intensity of the dynamic-magnetization pulse falls below 50% of its maximum value achieved with long pulses.

To estimate the shortest pulse achievable without an appreciable loss of intensity, we note that the actual temporal width of the generated pulse of the dynamic magnetization is smaller than that of the driving-current pulse. For example, Gaussian fitting of the magnetization pulse induced by the pulse of current with the width $w_d$ = 3 ns (Fig. 3(a)) yields a half-maximum width of 2.1 ns. Therefore, one can conclude that the technique allows one to excite pulses of the dynamic magnetization with the width down to about 2 ns, without significantly compromising the efficiency of the conversion of the dc current pulse into a microwave-frequency signal. Generation of shorter microwave pulses can also be achieved, but at the expense of the reduced power efficiency.

We now turn to the analysis of spin waves excited in the nano-waveguide by the current-induced auto-oscillations. To investigate the propagation of the excited spin waves, we applied driving-current pulses with the width $w_d$ = 3 ns and amplitude 7 mA, and mapped the time-dependent BLS intensity by rastering the probing laser spot over a 4.5 μm x 0.8 μm region encompassing the NLSI oscillator and the waveguide. The BLS signals were recorded at the frequency $f$=8.0 GHz of the auto-oscillations at this current (see Fig. 1(b)). The BLS maps, obtained at different delays relative to the start of the current pulse, are shown in Fig. 4(a). To better visualize the spatial characteristics of the propagating spin-wave packet, these maps are compensated for the spatial decay of spin waves by multiplying the experimental data by $\exp(2x/\xi)$. Here, $\xi$ is the spin-wave propagation length defined as the distance over which the wave amplitude decreases by a factor of $e$. The value $\xi$ =2.5 μm is determined based on the exponential fit of the propagation-coordinate dependence of the spin-wave intensity, integrated over the transverse direction of the maps and the propagation time of the packet (Fig. 4(b)).

The maps shown in Fig. 4(a) demonstrate that the pulse of the current-induced magnetization precession of the NLSI oscillator efficiently couples to spin waves in the waveguide, producing a propagating spin-wave packet. At $t$=1.6 ns, a dynamic signal emerges in the waveguide near its edge facing the NLSI oscillator. At $t$=2.4 ns, the increased-intensity region spreads away from the oscillator, indicating the propagation of the leading front of the spin-wave packet. Finally, at $t$=3.2 ns the packet occupies almost the entire analyzed length of the waveguide.

We note that the studied wave packet has a small temporal width of about 2.1 ns, but a large spatial width of more than 4 μm comparable to the size of the measured maps. The relation between these characteristics is determined by the large group velocity of spin waves in the waveguide, which is advantageous for the reduced spatial propagating losses, but complicates the analysis of the propagation characteristics in the spatial domain. Therefore, we concentrate on the time-domain analysis of the wave packet propagation (Fig. 5). We fit the temporal profiles of the spin-wave packet recorded at different distances from the NLSI oscillator by a Gaussian function (inset in Fig. 5(a)), and determine the dependence of the propagation delay (Fig. 5(a)) and the temporal width (Fig. 5(b)) of the packet on the propagation coordinate $x$ measured from the center of the oscillator along the waveguide. The propagation delay exhibits a well-defined linear dependence on $x$, enabling one to accurately determine the spin-wave group velocity $v_g$. From the linear fit of the data in Fig. 5(a), we obtain $v_g$=2.5 μm/ns, which is in a good agreement with the micromagnetic simulations performed in Ref. 18.

The dependence of the temporal width of the spin-wave packet on the propagation coordinate (Fig. 5(b)) reveals an intriguing behavior. Based on the general theory of waves in dispersive media, one can expect that the short wave packet should experience a temporal broadening due to the different velocities of its spectral components. Contrary to these expectations, the data of Fig. 5(b) demonstrate that the wave packet experiences a noticeable compression from the temporal width of 2.06±0.05 ns at the edge of the guide, to 1.73±0.05 ns at the distance of 1.5 μm from the edge. The initial compression is followed by a monotonic broadening at larger propagation distances. The only known mechanism that can be responsible for the observed temporal compression is the dynamic magnetic nonlinearity, which under certain conditions can counteract the dispersion broadening and lead to the formation of spin-wave solitons[33,34]. This interpretation is consistent with the observed dependence on the propagation distance: the compression is observed only at the initial stage of the packet propagation, where the amplitude of the dynamic magnetization in the spin wave is sufficiently large. As the amplitude of the propagating wave decreases due to damping, the nonlinear effects disappear, and the wave packet starts to broaden due to the expected effects of dispersion. We emphasize a significant potential of the observed nonlinear phenomena for applications, where they can be utilized to further reduce the width of the generated wave packets, improving the information transmission capacity of magnonic nano-circuits.

In conclusion, we have demonstrated the unique benefits of NLSI oscillators as nanocale sources of short spin-wave packets for the implementation of high-speed magnonic devices. Their intrinsically fast response to the driving electric current pulses enables efficient excitation of nanosecond spin-wave packets. In addition, the large amplitude of the excited magnetization dynamics allows one to implement nonlinear compression that can counteract the undesired dispersion spreading of short wave packets in magnonic nano-waveguides. We believe that our results should stimulate further developments in magnonics, and bring this area of research closer to the real-world applications.

This work was supported in part by the Deutsche Forschungsgemeinschaft, NSF Grants ECCS-1509794, DMR-1504449, and the program Megagrant № 14.Z50.31.0025 of the Russian Ministry of Education and Science.

**FIGURE CAPTIONS**

Fig. 1 (color online) (a) Schematic of the experiment. Inset illustrates the mechanism of pure spin current injection in the vertical NLSI spin valve. (b) BLS spectra of the current-induced magnetization oscillations at different currents, as labeled. Vertical dashed line marks the frequency $f_0$ of the uniform ferromagnetic resonance in the Py film. The linewidth of the spectral peaks is determined by the limited spectral resolution of the BLS setup. The data were obtained at $H_0 = 1000$ Oe.

Fig. 2 (color online) (a) Time dependence of the auto-oscillation intensity recorded with 100 ns-wide pulses of driving current, at the labeled current amplitudes. (b) Leading edge of the auto-oscillation intensity pulse on the log-linear scale. Lines show the exponential fit of the experimental data. (c) Current dependence of the rise time of the auto-oscillation intensity. Curve is a guide for the eye. The data were obtained at $H_0 = 1000$ Oe.

Fig. 3 (color online) (a) Time dependencies of the auto-oscillation intensity recorded by applying pulses of the driving current with the amplitude of 7 mA and different widths, as labeled. The data for the 3-ns wide pulse are fitted by the Gaussian function. (b) Dependence of the peak intensity of the auto-oscillation pulse on the width of the driving-current pulse. Curve is a guide for the eye. The data were obtained at $H_0 = 1000$ Oe. The auto-oscillation frequency is 8 GHz.

Fig. 4 (color online) (a) Normalized decay-compensated maps of the spin-wave intensity recorded at delays of 1.6, 2.4 and 3.2 ns with respect to the start of the driving current pulse, as labeled. Dashed lines indicate the contour of the nano-waveguide. (b) Propagation-coordinate dependence of the spin-wave intensity integrated over the transverse sections of the maps and over the propagation time, on the log-linear scale. Solid line is the exponential fit of the experimental data. Vertical dashed line marks the position of the nanocontact. The data were obtained at $H_0$ = 1000 Oe. The width of the driving-current pulse is 3 ns and its amplitude is 7 mA. The spin-wave frequency is 8 GHz.

Fig. 5 (color online) Propagation-coordinate dependence of the propagation delay (a) and of the temporal width (b) of the spin-wave packet. Solid line in (a) is a linear fit of the experimental data. Inset shows the temporal profile of the wave packet at $x$=0, fitted by the Gaussian function. The data were obtained at $H_0$ = 1000 Oe. The width of the driving-current pulse is 3 ns and its amplitude is 7 mA. The spin-wave frequency is 8 GHz.

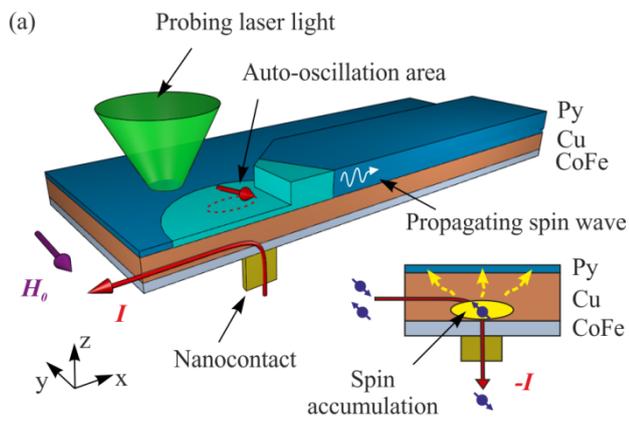

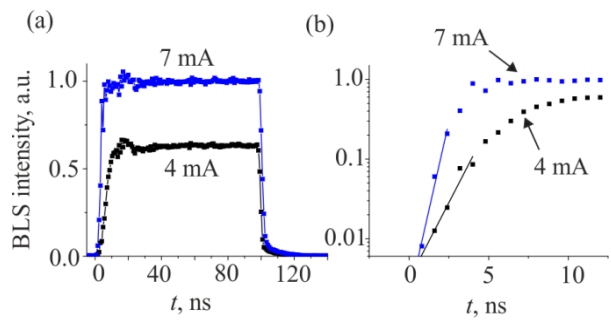

Fig. 1

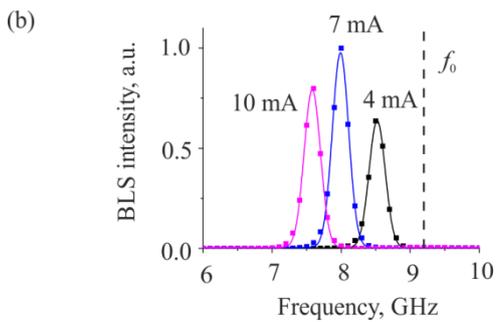

Fig. 2

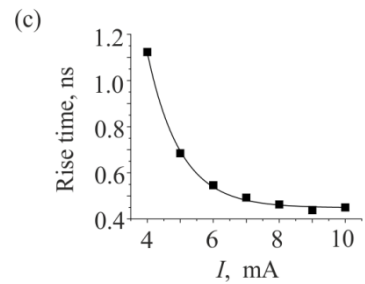

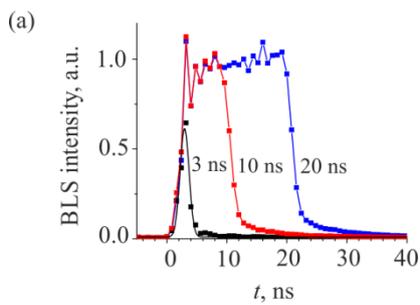

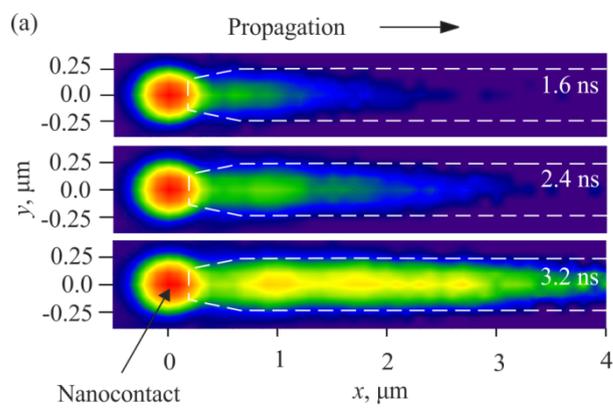

Fig. 3

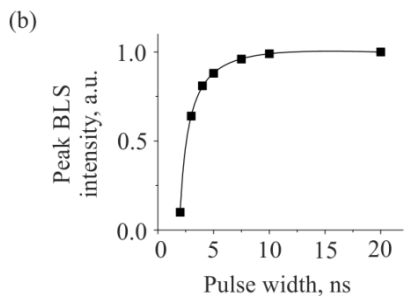

Fig. 4

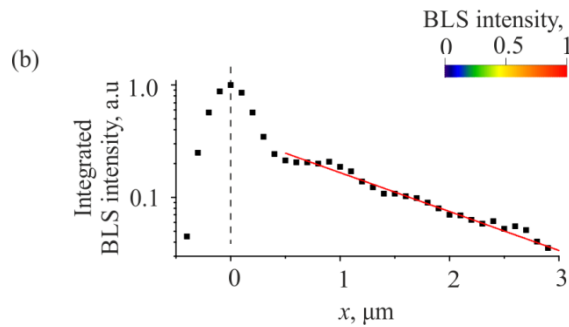

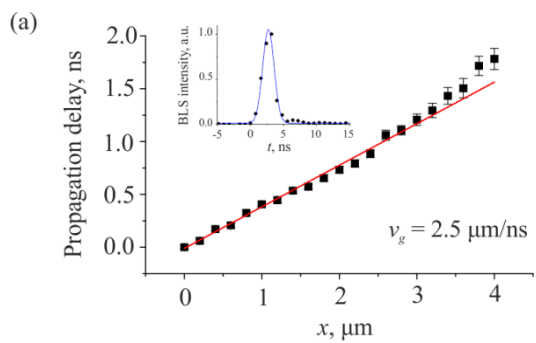

Fig. 5

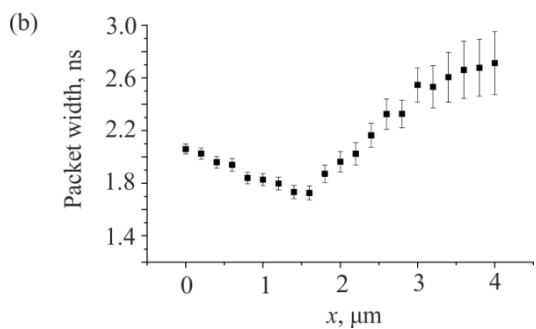